# Microchannel-Plate Detector Development for Ultraviolet Missions

Lauro Conti[a*] , Jürgen Barnstedt[a], Sebastian Buntrock[a], Sebastian Diebold[a], Lars Hanke[a], Christoph Kalkuhl[a], Norbert Kappelmann[a], Thomas Kaufmann[a], Thomas Rauch[a], Beate Stelzer[a], Thomas Schanz[a], Klaus Werner[a], Hans-Rudolf Elsener[b], Sarah Bougueroua[c], Thomas Keilig[c], Alfred Krabbe[c], Philipp Maier[c], Andreas Pahler[c], Mahsa Taheran[c], Jürgen Wolf[c], Kevin Meyer[d], Daniel M. Schaadt[d]

[a]Institute for Astronomy and Astrophysics Tübingen, Kepler Center for Astro and Particle Physics, University of Tübingen, Sand 1, 72076 Tübingen, Germany
[b]Empa, Swiss Federal Laboratories for Materials Science and Technology, Ueberlandstrasse 129, Dübendorf, Switzerland
[c]Institute of Space Systems, University of Stuttgart, Pfaffenwaldring 29, 70569 Stuttgart, Germany
[d]Institute of Energy Research and Physical Technologies, Clausthal University of Technology, Leibnizstraße 4, Clausthal-Zellerfeld, Germany

## ABSTRACT

The Institute for Astronomy and Astrophysics in Tübingen (IAAT) has a long-term experience in developing and building space-qualified imaging and photon counting microchannel-plate (MCP) detectors, which are sensitive in the ultraviolet wavelength range. Our goal is to achieve high quantum efficiency and spatial resolution, while maintaining solar blindness and low-noise characteristics.

Our flexible detector design is currently tailored to the specific needs of three missions: For the ESBO *DS* (European Stratospheric Balloon Observatory – *Design Study*) we provide a sealed detector to the STUDIO instrument (Stratospheric Ultraviolet Demonstrator of an Imaging Observatory), a 50 cm telescope with a UV imager for operation at an altitude of 37-41 km. In collaboration with the Indian Institute of Astrophysics we plan a space mission with a CubeSat-sized far-ultraviolet spectroscopic imaging instrument, featuring an open version of our detector. A Chinese mission, led by the Purple Mountain Observatory, comprises a multi-channel imager using open and sealed detector versions.

Our MCP detector has a cesium activated p-doped gallium-nitride photocathode. Other photocathode materials like cesium-telluride or potassium-bromide could be used as an alternative. For the sealed version, the photocathode is operated in semi-transparent mode on a $MgF_2$ window with a cut-off wavelength of about 118 nm. For missions requiring sensitivity below this cut-off, we are planning an open version. We employ a coplanar cross-strip anode and advanced low-power readout electronics with a 128-channel charge-amplifier chip.

This publication focuses on the progress concerning the main development challenges: the optimization of the photocathode parameters and the sophisticated detector electronics.

## 1. INTRODUCTION

We develop MCP detectors suitable for imaging and spectroscopy in the ultraviolet (UV) which can be adjusted to the needs of specific missions. Such detectors are available for scientific purposes since the late 1970s, a comprehensive overview about the topic can be found in Wiza (1979)[1]. The working principle of such a detector is as follows: an incident UV photon produces a photoelectron in a suitable photocathode. The conversion probability, the so-called quantum efficiency (QE), is wavelength dependent. The photoelectron can then be multiplied in a stack of two MCPs in chevron configuration and the resulting charge cloud is detected on a position sensitive anode. The front-end electronics analyze the charge information from the anode to determine the position of the incident photon. Such a detector is photon counting and solar blind by design – as photocathode materials with bandgaps above about 3.1-4.1 eV are used, which corresponds roughly to 400-300 nm. There are two different modes for the photocathode. In semi-transparent mode the photocathode is grown on a UV transparent window of the detector, in opaque mode on the surface of the first MCP, for an overview see Spicer and Herrera-Gomes (1993)[2]. With $Cs_2Te$ as photocathode material 5-10 % QE can be achieved in semi-transparent mode on quartz (see Sect. 3.1). With p-doped GaN in opaque mode very high QEs of 50 %, at 250 nm, and even higher values for shorter wavelengths were obtained by Siegmund et al. (2017)[3].

---

[*]conti@astro.uni-tuebingen.de

MCPs and photocathodes need to be operated in vacuum. Either the detector is sealed with a UV transparent window (see Sect. 2.3.1) or a door mechanism is opened, once the detector is in an ultra-high vacuum (UHV) for testing or finally in space. The wavelength range for the sealed version is determined by the cutoff wavelength of the window material, with about 113 nm for $MgF_2$ or 105 nm for LiF at the short wavelength end, as reported by Hunter et al. (1969)[4]. For the detector variant with a door instead of a window, bare atomic layer deposition (ALD) MCPs yield acceptable QEs of about 10 to 20 percent between 100 nm to 40 nm, but would increase in the entire UV wavelength range e.g. with a CsI photocathode on the first MCP surface.

For our detector, the spatial resolution is about 2000 × 2000 pixels on an active image area of 39 mm × 39 mm, i.e. a pixel size of 20 µm. Tests with our 64 + 64 cross-strip (CS) readout indicate that even higher resolutions could be achieved, e.g. with MCPs featuring smaller capillary pitch, which is 13 µm for our current ALD MCPs from Incom.

The bottleneck for the count rate of our current detector are the four ADCs which convert the analogue signals of our charge amplifier chip with a maximum global count rate of 300,000 counts/s. The local count rate, which would be important for bright point sources when imaging, is limited by the recharge time of the MCP channels and is as high as 40 to 80 counts per pixel and second.

A major advantage of our MCP detector compared to CCDs and CMOS sensors is that it has no readout noise and only low dark noise with $2.5 \times 10^{-4}$ counts per second and pixel without the need for cooling. Further improvements in dark noise have been achieved by switching from lead glass to borosilicate glass as MCP substrate, since the dark noise due to the decay of radioactive isotopes is comparatively high in lead glass.

For a quick overview of our detector, Table 1 presents some of the capabilities and most remarkable characteristics.

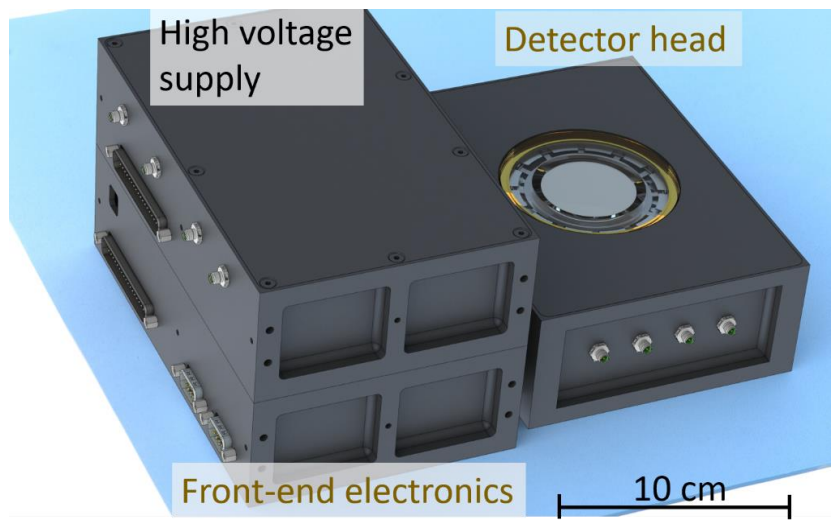

Figure 1. Detector assembly without cables and mounting brackets. The position of the detector head can be adjusted with a maximum distance of about 50 cm from the FEE.

Table 1: Specifications and properties of the IAAT detector versions.

|  | **Closed IAAT detector** | **Open IAAT detector** |
|---|---|---|
| Sealing | (0 0 1) $MgF_2$ window, coated sealing surfaces with good wetting properties for molten indium alloy | Simple and lightweight door mechanism |
| Photocathode | $Cs_2Te$ and cubic GaN on $MgF_2$ with depth graded magnesium (p-doping) | GaN, other photocathode materials possible |
| MCPs | Temperature resistant to about 450 °C for getter activation, ALD-MCPs with long lifetime, 13 µm capillary pitch and small gain modulations | Like closed version. NiCr as electrode material or buried electrode and MgO as top layer |
| Mass | FEE, HV supply, detector head and cables about 3 kg | Like closed detector, additionally less than 1 kg for door mechanism |
| Anode | Coplanar CS with 64 strips per axis, 39 mm × 39 mm. Low Temperature Cofired Ceramics (LTCC) is used as substrate, High Temperature Cofired Ceramics (HTCC) version is possible | |
| Preamplification | Radiation hard 128-channel 40 MHz BEETLE chip with equivalent noise charge of only about 500 electrons for low load capacitances | |
| Front end electronics | Xilinx Virtex-5 board with four ADCs, receiving the BEETLE signals and I²C interface to control the BEETLE chip. Flight version with space-grade components possible. Centroiding algorithms in hardware, 4k image is evaluated, 2k × 2k image is stored on SRAM | |

## 2. DETECTOR CONCEPT

### 2.1 Detector parts

The detector consists of three boxes (see Figure 1): the detector head, the front-end electronics including the power board, and the high voltage supply. The total mass including cables is about 3 kg.

The central component of the detector is the detector head. In semi-transparent mode, a voltage between the photocathode and surface of the first MCP is applied. Electrons leaving the photocathode will have a certain velocity parallel to the MCP surface, increasing the achievable resolution for smaller distances and higher voltages, as described by Csorba (1977)[5]. In our case we plan that this resolution-reducing distance to the first MCP is only 200 to 300 µm, hence the tolerance for the height of the detector body is only 100 µm. As MCPs we have chosen ALD-MCPs with MgO secondary electron emissive (SEE) material. Due to the combination of a higher lifetime of these MCPs compared to conventional types, i.e. extractable charge per area and less ion feedback, the lifetime of the whole detector is highly increased. Furthermore, ALD-MCPs can withstand temperatures above 450°C, allowing the integration and activation of a getter inside the detector housing for an even longer detector lifetime.

The laboratory version, balloon version and space version of our high voltage power supply have been successfully tested and used. For one of the applications of our detector, the ESBO DS stratospheric balloon flight (see Chap. 6), we need to operate the HV supply in a height of 40 km with an atmospheric pressure of about 1 mbar. To prevent accidental arcing, we employ molded electronics and use only one sophisticated adapter plug to facilitate the assembly and disassembly of the HV supply and detector box.

For the FEE we designed a custom board with a Xilinx Virtex-5 FPGA. Currently, we test all components on the board. As a next step we will implement promising algorithms in hardware and compare their results to those we achieve in software (for the latter see Sect. 5.2). Using a Virtex-5 FPGA has several advantages: there is a space qualified version available and it has enough pins to contact the SRAM and the four ADCs for the input data stream. The SRAM is required to store lookup tables and for the image integration. It also has enough digital signal processing (DSP) slices, for the algorithms to run fast and energy-efficiently. For each charge data sample, we need only a few clock cycles for our calculations. As shown in Figure 2, the FEE houses several boards. Besides the FPGA board, a separate SRAM board is used, which allows the FEE to be kept more compact and is advantageous when creating a new version of the FPGA board, e.g.

with different connectors. The DC-DC converters on the power board in the FEE are electronically isolated and transform the nominal input voltage of 27 V (24-36 V) to the voltages required by all components of the detector system (but the high voltage). The flight version of the FEE looks like the laboratory version but uses more advanced cooling solutions, e.g. copper plates, heat pipes, and vapor cambers to cool the electronics. The total power budget of the detector when imaging is 15 W. Every component of the FEE is designed to also work in vacuum or under stratospheric conditions. To simplify the implementation of our detector to different instrument platforms, we are using a single power connector and data connector (plus redundancy) for the whole detector system. For further details on the FEE including the electronic schematics see Sect. 5.1.

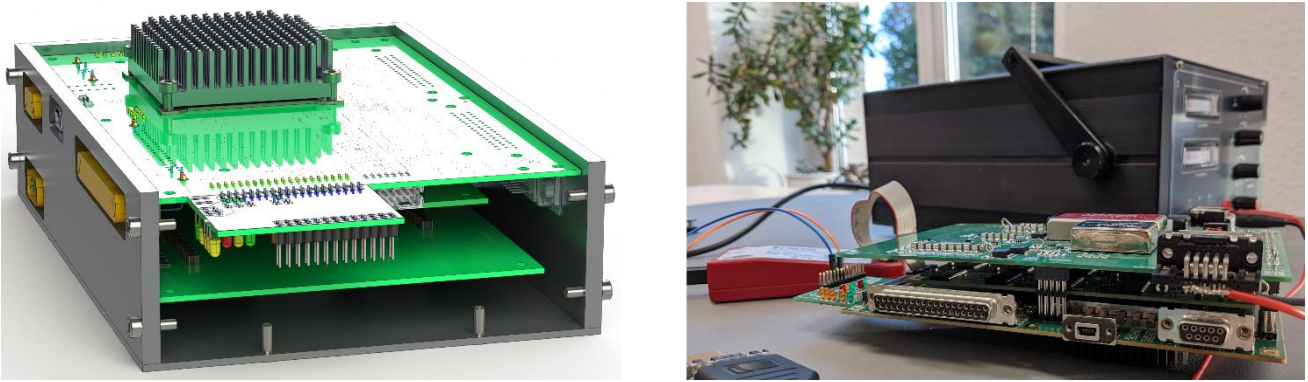

Figure 2. FEE with the Virtex-5 FPGA board, SRAM board, and power board. On the laboratory versions shown here, a passive air cooler removes the heat from the FPGA. A debug board in the front of the laboratory and flight versions of the FEE can be removed easily, e.g. before flight. Commands can be sent to the HV supply via a 37-pin D-Sub connector. The flight version for the ESBO *DS* mission uses a USB 2.0 port with a screwable mini B plug for communication with the instrument computer. In the CAD drawing shown on the left, two side panels were hidden to show the boards inside of the FEE. With the help of such a CAD drawing suitable housings and cooling solutions for the different FEE versions can be easily planned and tested.

## 2.2 Detection chain

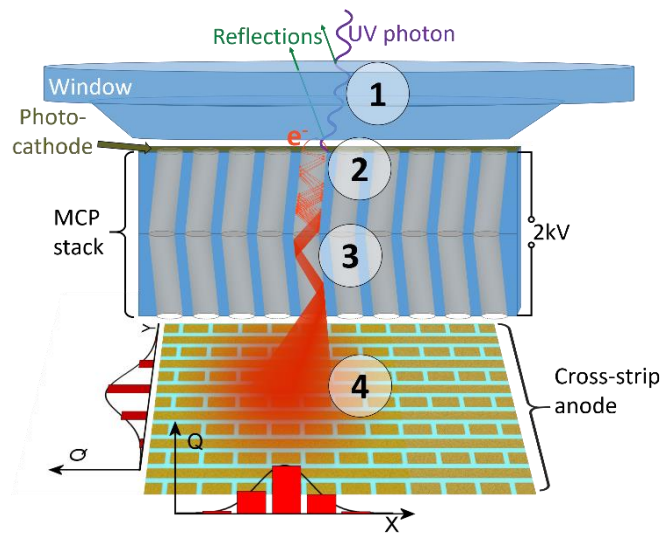

Figure 3. Detection chain of the UV MCP detector, the individual parts are not to scale. Note the QX-diagram at the bottom and QY-diagram on to bottom left plotting the actual charge deposited on corresponding anode strips with a Gaussian fit for illustration purposes.

An image or spectrum on the focal plane consists of individual photons. The goal of our detector is to determine the position and arrival time of as many of these photons as possible with an accuracy of slightly less than 20 µm. To gain color information about an object, filters, a spectrograph, or an imaging spectrograph can be used.

As the detector is photon counting, it is sufficient to understand the detection of a single photon, the process is then repeated up to a few hundred thousand times per second when operating the detector.

With reference to the numbers in Figure 3, we now go through the detection chain, this time focusing on the main properties of the parts that are involved in these steps and their main characteristics that influence the performance of the detector:

1. The UV photon passes through the $MgF_2$ window. The reflection probability is low if the photon incidents almost perpendicular to the window surface, since the refractive index of $MgF_2$ is approximately one in the relevant wavelength range. The reflected intensity peaks at a wavelength slightly below the cutoff of $MgF_2$ at 118 nm with about 6 %, as calculated with data from Rodríguez-de Marcos et al. (2017)[6]. Furthermore, the opacity of the window is almost vanishing low wavelength larger than the cut-off (see Sect. 2.3.1). Birefringence is avoided by using (0 0 1) $MgF_2$. An open version of the detector mitigates reflections, absorption and especially the short wavelength cut-off, but a door mechanism is required to protect the detector under atmospheric conditions.

2. The UV photon is absorbed by the photocathode and a photoelectron is generated. The typical absorption length of suitable photocathode materials should be of course much smaller than the photocathode thickness. Further details on the properties of our photocathodes are discussed in Chap. 3.

3. The photoelectron is accelerated towards the MCP. The MCP has an open area ratio (OAR), i.e. the capillary area with its high secondary electron emissive (SEE) walls to the total area of the electrode, which is larger than 65 % for the ALD-MCPs we are using. Photoelectrons hitting the SEE walls produce about two to seven secondary per primary electrons, depending on the material used and its thickness, as reported by Mane et al. (2012)[7]. We are applying voltages of about 2 kV between the surface of the first MCP and the bottom of a second MCP, with a resulting electron avalanche consisting of about $10^5$ electrons at the bottom of the second MCP.

4. The anode comprises 64 + 64 individual strips in x- and y-direction and is embedded in electrically insulating low temperature cofired ceramics (LTCC). We use a coplanar design, i.e. the strips in x-direction are completely buried under the LTCC, rectangular vias lead to the surface and end in the same plane as the strips in y-direction. The charge is deposited on the anode within less than about 100 ns.

In the preamplifier chip the charge deposited on each of the 128 strips is sampled with 40 MHz into a ring buffer. If a trigger occurs, which is defined by a set of adjustable parameters, the respective samples are multiplexed to four outputs. This approach saves resources, as our FEE must only receive, convert, and analyze real events. In the FEE, the charge information is analyzed with a centroiding algorithm (see Sect. 5.2) to determine the center of mass of the electron cloud. The detector has two image modes: In the photon-by-photon mode, position and timestamp are stored for each event; in the image-integration mode, a 2k × 2k greyscale image is produced, the greyscale value of each pixel encodes the number of events on a corresponding pixel position.

## 2.3 Detector variants

### 2.3.1 Sealed, closed detector design

The MCPs and especially the cesium activated photocathodes must be kept under ultra-high vacuum (UHV), e.g. good results when activating photocathodes with cesium can be achieved, when the partial oxygen pressure is less than $5 \times 10^{-12}$ mbar. For detailed information about our vacuum chambers see Hermanutz et al. (2014)[8]. The (0 0 1) $MgF_2$ window we use for sealing has a transmittance of over 90% for wavelengths below 170 nm and a cut-off, i.e. the transmittance lower than 66 %, at 118 nm. The window thickness is 9 mm, in the middle of the window an area with a diameter of 41 mm can be used to grow a photocathode on the window if the detector is to be operated in semi-transparent mode. For photocathodes with high sheet resistance it is necessary to first deposit a UV-transparent electrode on this area, for details see Sect. 3.1. Sect. 4.1 describes the main challenges we solved to successfully seal the detector.

### 2.3.2 Open detector design

If measurements below the cut-off of suitable UV windows are required, a door or valve can be deployed for sealing. In this case the photocathode is grown on the first MCP and is operated in opaque mode. For the growth of GaN, substrate temperatures of at least 450°C are required, as described in Meyer et al. (2019)[9]. Currently we use mechanical ALD-MCPs samples and similar suitable samples from Incom, Photonis, and CrysTec, which withstand these temperatures and optimize the growth parameters of GaN on these substrates.

## 3. PHOTOCATHODE

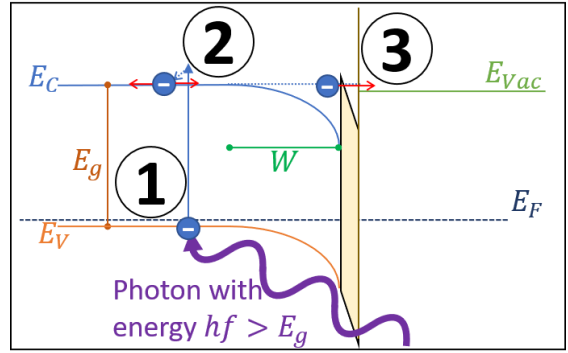

Figure 4. Band structure and three-step model in case of a p-doped NEA photocathode in opaque mode with a layer of Cs on the surface (yellow area). Explanations and definition of the physical quantities can be found in the text.

Using the three-step model of photoemission by Spicer (1958)[10], the quantum efficiency of the photocathode can be expressed as the product of the corresponding probability of each step to occur. Referring to Figure 4 these probabilities and possible measures to increase them are:

1. The probability for the absorption of UV photons by the photocathode. An electron on the valence band with a potential energy $E_V$ can be excited to the conduction band with energy $E_C$ by absorbing a photon with an energy of at least the bandgap $E_C - E_V = E_g$ of the crystal. Photons have an energy dependent absorption length, i.e. an absorption probability of 63 %. We are using a telluride thickness of 12 nm for $Cs_2Te$ photocathodes and at least 100 nm for GaN photocathodes in semi-transparent mode. In semi-transparent mode, the thickness of the photocathode is a compromise between the absorption probability of photons and their scattering length, which is described in the following point.

2. The probability for electrons to reach the surface with a potential energy slightly above the conduction band edge. We use only low electric fields, so that the electrons remain "slow", i.e. their diffusion length can be in the order of micrometers with accordingly very low losses. For high electric fields we observe a negligible increase in the total quantum efficiency but a rising dark current. The p-doping of the photocathode can introduce scattering centers in the photocathode crystal, decreasing the diffusion length.

3. The probability for the electrons on the conduction band to tunnel into the vacuum. For our photocathodes, the photocathode should have a negative electron affinity (NEA), i.e. the potential energy of the electrons on the conduction band in the bulk of the photocathode is above the vacuum level $E_{Vac}$. To achieve the NEA, in the case of GaN photocathodes we use two main measures: the cesiation of the surface of the photocathode and p-doping. As shown in Figure 4, the electron can tunnel through a thin remaining potential barrier of the cesium activation layer into the vacuum. Due to p-doping the energy $E_C$ in the bulk can be bigger than $E_{Vac}$. Only if the electron loses its energy within the depletion width $W$ due to scattering, it could be trapped in the so-called bent-band region and recombines after some time. A decreasing quantum yield is possible for high energetic photons in opaque mode, if they have a very small absorption length i.e. are likely to be absorbed within the depletion width $W$. For further details see the publications of Marini et al. (2018)[11] and Aslam et al. (2005)[12].

### 3.1 Cesium-telluride photocathode

We have optimized the growth of cesium-telluride photocathodes in semi-transparent mode. Due to a lack of conductivity of the photocathode material itself, the photocathode could become positively charged when illuminated, causing the effective electric field between the detector window and the first MCP to decrease, which in turn leads to a lower and not exactly defined detector efficiency. When using the detector for photometry, a slightly lower detector efficiency would be acceptable if, in return, it is stable for different illumination intensities. For this reason, we use thin Ni or NiCr layers to ground the photocathode in semi-transparent mode. To achieve an optimal contact we used 5 nm of NiCr, as it is expected by Zhou et al. (2008)[13] to form a continuous layer and have a low sheet resistance per film thickness and a transmittance of 25 to 19 %. Despite the high absorption of the metallization layer we achieved a detection efficiency of 3 % at 280 nm in semi-transparent mode with perfect electrical contact, with a corresponding internal QE of the photocathode of about 10 %. cesium telluride forms different alloys, we use $Cs_5Te_2$ for the highest response at about 285 nm and $Cs_2Te$ with a peak at lower wavelengths. For next batches we will apply 3.5 nm thick layers of pure Ni on the substrate, as it is also very stable against air and could have only a slightly lower sheet resistance despite a lower thickness.

### 3.2 GaN photocathode

The growth of the gallium nitride photocathode is performed at the Institute of Energy Research and Physical Technologies (IEPT). A Riber Compact 21 molecular beam epitaxy (MBE) system in combination with an Oxford Research plasma cell is used for the growth of the GaN thin film. To get information about the crystallographic structure of the thin film, the high-resolution x-ray diffraction system (HRXRD) Discover D8 from Bruker AXS is used.

For the ESBO *DS* project, the detector is operated in semi-transparent mode. Therefore, the GaN thin film is grown directly on a $MgF_2$ substrate. $MgF_2$ has a rutile crystal structure with lattice parameters of a = 0.462 nm and c = 0.305 nm. Due to these parameters it has a much smaller mismatch to cubic or hexagonal GaN than e.g. sapphire. Furthermore, it should be possible to grow both crystal structures of GaN on $MgF_2$, whereas on sapphire only hexagonal GaN can be grown. The lattice mismatch between GaN and $MgF_2$ is approximately in the range of 2.5 % to 4.5 %, dependent on the crystal structure of GaN.

In the first studies, the possibility of the direct growth on $MgF_2$ was investigated, more details are presented in two publications by Meyer et al. (2020)[9,14]. Three different crystallographic orientations of the substrate, (1 0 0), (0 0 1) and (1 1 0) were tested and afterwards analyzed by HRXRD. The ω-2θ scans of the three substrate orientations show different results, which are summarized in Table 2.

Table 2: Summary of the HR-XRD results of the GaN thin film on different $MgF_2$ substrate orientation in a growth temperature range of 450°C to 600°C: On (1 0 0) and (0 0 1), the thin film consists mainly of hexagonal GaN with cubic inclusions. On a (1 1 0) substrate, a mixture of both crystal structures can be measured.

| substrate orientation | hexagonal GaN reflexes | cubic GaN reflexes | thin film |
| --- | --- | --- | --- |
| $MgF_2$ (1 0 0) | (0 0 0 2) and (0 0 0 4) | (2 2 0) | mainly hexagonal with cubic inclusions |
| $MgF_2$ (0 0 1) | (0 0 0 2) and (0 0 0 4) | (2 2 0) | mainly hexagonal with cubic inclusions |
| $MgF_2$ (1 1 0) | (0 0 0 2) and (0 0 0 4) | (2 0 0) and (4 0 0) | mixture of hexagonal and cubic |

On all three orientations of $MgF_2$, a growth of a GaN thin film is possible. Both crystal structures of GaN can be detected but their crystallographic orientations are different depending on the orientation of the substrate. On an (0 0 1) and (1 0 0) substrate, mainly hexagonal (0 0 0 1) GaN is grown with some inclusions of cubic (1 1 0) GaN, whereas on (1 1 0) $MgF_2$, the thin film consists of a mixture of (0 0 0 1) hexagonal GaN and (1 0 0) cubic GaN. These results are independent of the growth temperature in a range of 450°C to 600°C. At growth temperatures up to 700°C, the layer quality is much worse, and the cubic part of the thin film disappears. The studies showed that a growth of GaN directly on $MgF_2$ is possible for three different crystallographic orientations in a temperature range of 450°C up to 600°C. First magnesium doped p-GaN thin films were grown, showing no differences in the crystallographic behavior to the undoped GaN thin films.

For future missions, an open-detector concept is planned. Therefore, first studies of the growth of GaN directly on MgO were performed (and will be published soon). Powell et al. (1990)[15], (1993)[16] showed previously the growth of GaN directly on MgO. Due to the cubic crystal structure of MgO, mainly cubic GaN is grown with inclusions of hexagonal GaN. The results of the study at the IEPT are summarized in Table 3.

Table 3. Summary of the HR-XRD results of the GaN thin film on MgO under different growth conditions: Under N-rich conditions, a mainly cubic GaN thin film with hexagonal inclusions can be detected. Under Ga-rich conditions, only cubic (1 0 0) GaN is grown.

| growth conditions | hexagonal GaN reflexes | cubic GaN reflexes | thin film |
|---|---|---|---|
| N-rich | (0 0 0 2) and (0 0 0 4) | (2 0 0) and (4 0 0) | mainly cubic with hexagonal inclusions |
| Ga-rich | - | (2 0 0) and (4 0 0) | only cubic |

The growing process was performed at 600°C. The samples were analyzed by HRXRD afterwards like the MgF$_2$ samples above. The GaN thin film consists of mainly cubic GaN with inclusions of hexagonal GaN under rather N-rich conditions, comparable to Powell et al. (1990)[15]. Under Ga-rich conditions, the thin film consists only of cubic (1 0 0) GaN. At these conditions, the formation of Ga droplets on the surface can be observed, which can be removed by hydrochloric acid. More studies on this topic are planned, like the doping with Mg and the activation with cesium to increase the quantum efficiency, comparable to the studies of GaN on MgF$_2$. The first results demonstrated the high potential of the direct growth of GaN on MgO. Further results of the studies are planned to be published in the future.

## 4. DETECTOR SEALING

### 4.1 Optimization of detector sealing

In Sect. 2.3.1 the sealed detector variant is described. Successful sealing of the detector, i.e. the cesium containing photo-cathode does not degrade within several weeks even when the detector body is heated and cooled in the required temperature range in humid air, demands a combination of optimization steps. In our case, the suboptimal wetting of the sealing surfaces had the greatest influence. We solved the problem by comparing different tests with promising material combinations in a drop shape analyzer and coating our sealing surfaces accordingly.

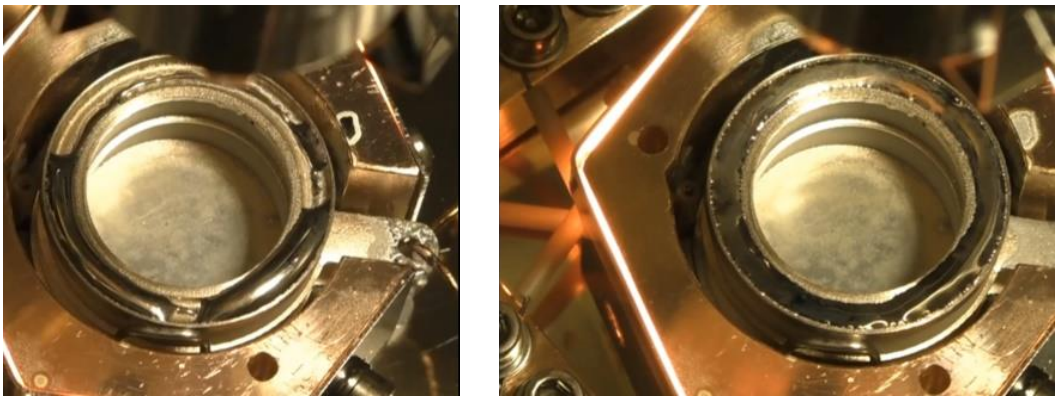

Figure 5. Diode body for sealing tests of 2.50 cm samples in vacuum with molten indium-alloy. Measurements of a temperature sensor next to the diode body are used for calibration purposes. On the left shortly after the indium melts. On the right, due to the low surface tension between the coated sealing area and the alloy, it spreads evenly over the entire area within hours.

### 4.2 Changes in the properties of photocathodes during and after sealing

During the sealing process, the sealing areas near the cesium containing photocathode are heated to temperatures of up to 360°C. One problem to be solved is whether and to what extent cesium is evaporated or sublimated during the entire process. We have also found that, even with sensitive measurements, cesium telluride is stable at these temperatures, i.e., there is no measurable decrease in photocurrent. It seems that the chemical bond of the $Cs_2Te$ crystal is therefore high enough to withstand the sealing temperatures. In the case of cesium activated GaN, only a cesium activation layer is applied to the surface of GaN. It is expected that the first cesium layer is strongly bound on GaN, but even at room temperature other cesium layers sublimate slowly if the cesium partial vapor pressure within the volume to which the photocathode is exposed is too low. This is an expected behavior as we have measured it ourselves and as reported by Stock et al. (2005)[17]. Again, our measurements shortly before and after sealing show that at the given temperature and duration of the sealing process in our UHV system there is no major change in the response of the photocathode. More precise tests on GaN photocathodes are in progress to quantify the small losses that can nevertheless be expected.

After sealing, we mount our diode bodies into a holder with an integrated calibrated NIST photodiode. Using a monochromator, we can select narrow bands from the spectrum of a deuterium lamp and measure the corresponding photocurrent. We perform these measurements in a closed, light-tight and air-tight box, which we flood with a predefined amount of nitrogen to achieve reproducible conditions. Figure 6 shows that the quantum efficiency of the photocathodes in the sealed diode bodies did not change at all within weeks and even months, with the only deviations visible due to a worse SNR near the cut-offs of the quartz substrate or photocathode.

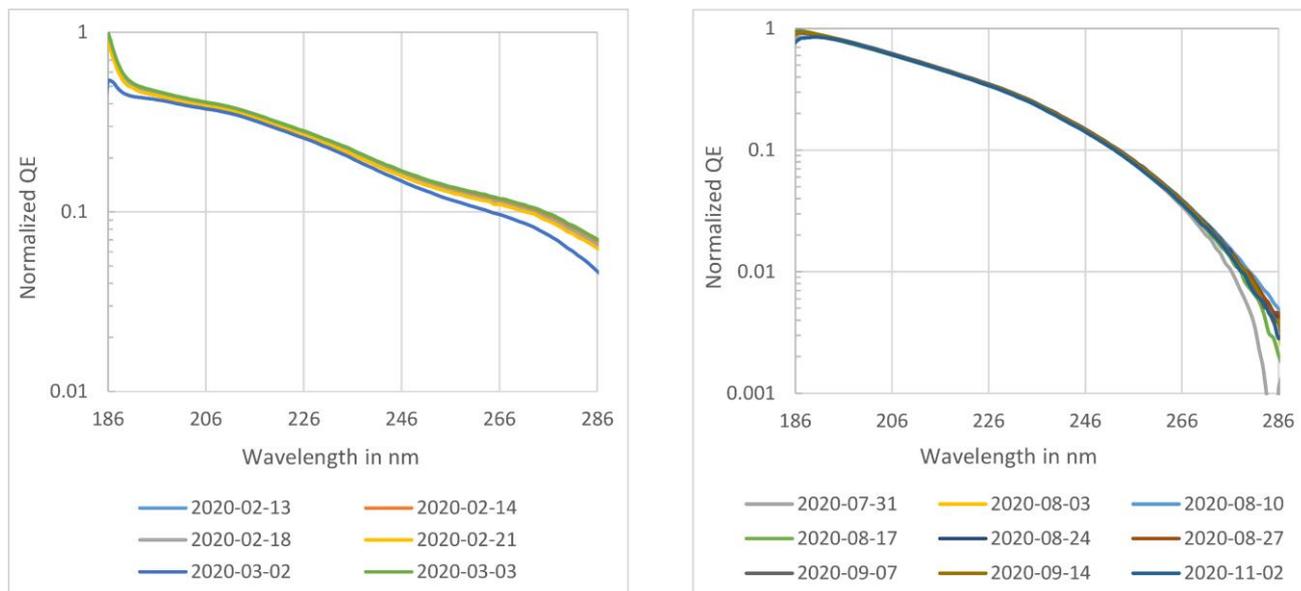

Figure 6. QE measurements over the wavelength on two sealed diode bodies with cesium telluride photocathode over a period of 2 weeks (left) or several months (right). The outlier on 2020-03-02 was measured after some tests in a climate chamber without significant deviation when re-measured the following day.

## 5. DATA PROCESSING

### 5.1 Detector electronics

Most parts of our detector electronics have been designed to take advantage of the capabilities of the 128-channel charge amplifying BEETLE chip. Details on its application can be found in previous articles by Pfeiffer et al. (2014)[18] and Conti et al. (2018)[19]. We use four ADCs to digitize the analog signals from the BEETLE chip and process the data with a centroiding algorithm implemented in an FPGA (s. following Section 5.2). A schematic of the electronics for the balloon version of our detector are shown in Figure 7.

We are currently testing and configuring the laboratory version of our front-end electronics with a commercial Virtex-5 FPGA and commercial grade components. For the balloon version, the FPGA will be exchanged for its industrial version, while for space missions a space-qualified version of the Virtex-5 is available.

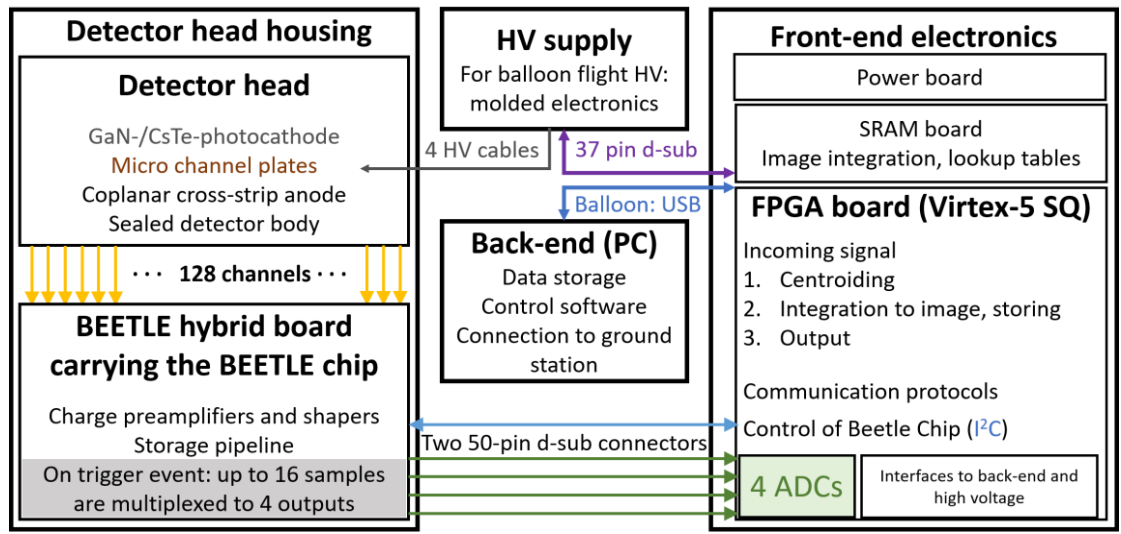

Figure 7. Schematic diagram of the detector electronics as used for the STUDIO instrument. For space missions, only the front-end electronics needs to be adapted, i.e. with suitable connectors to the back-end electronics and space-qualified versions of components like the Virtex-5 FPGA and the ADCs.

## 5.2 Centroiding methods

Each MCP pulse is distributed over about five anode strips in both x- and y-direction. The wings of the charge distribution may extend even further. To determine the position of the incoming event an algorithm is needed, which calculates the centroid of the charge position. As we have 64 strips in both directions, the position must be determined to ~1/32 strip pitch to reach the baseline resolution of 2000 pixels per axis or to ~1/64 strip pitch for achieving the goal of 4000 pixels per axis.

We evaluated several centroiding algorithms and estimated their accuracy and feasibility in an FPGA environment. Particular benchmarks were linearity, resolution, and possible artifacts within the image. We found that fitting a Gaussian to the charge distribution of each axis yields the best results with respect to resolution and linearity and produced almost no artifacts. However, this fitting method has a high demand on computing resources and the time for convergence is not determinate. We used the Gaussian fit as reference for determining the characteristics of other algorithms under evaluation. Therefore, we developed several tools for determining these characteristics. The ideal algorithm for the implementation in an FPGA uses simple integer arithmetic without iteration.

The best candidate to fulfill these requirements was found in modified method originally developed by Caruana et al., (1986)[20], and later improved by Guo (2011)[21]. It is based on the logarithm of the Gaussian function and results in a quadratic equation that can be solved analytically. The necessary calculation of the logarithm of the data can be simplified by using a lookup table and with an optimized normalization the calculation is possible in integer arithmetic. Our tests showed a performance quite close to the Gaussian fit method but with considerably less resource consumption.

Simple center of gravity algorithms were also evaluated and usually show a pronounced nonlinearity together with artifacts. Such a behavior was already described by Vallerga et al. (2011)[22], for their "all above threshold" (AAT) algorithm.

We also tested other more promising algorithms, which we call "pulse template method" (PTM) and in a modified version "template matching method" (TMM). These methods are based on lookup tables that contain normalized averaged pulse templates, for which the centroid was estimated by a Gaussian fit. For an individual event, the data are normalized and then correlated with the templates. The best fit is determined by a least squares algorithm. While the PTM method still produces some artifacts, these artifacts could be widely eliminated by the TMM. A detailed publication is in preparation.

In general, all methods produce a certain degree of nonlinearity. This nonlinearity manifests as intensity variations that are periodic with the strip pitch of the anode and can be determined via flat-field images. A correction of this effect is planned to be implemented as a lookup table.

## 6. CURRENT MISSIONS

During the H2020 funded European Stratospheric Balloon Observatory – *Design Study* (ESBO *DS*), a European consortium is developing a 50 cm aperture balloon-based observatory named STUDIO (Stratospheric UV Demonstrator of an Imaging Observatory). Furthermore, a roadmap towards larger aperture telescopes and towards an observatory-style balloon observatory is being developed. The first flight of the STUDIO flight system is foreseen for 2022 from Esrange in northern Sweden.

One of the goals of ESBO *DS* is to ease the access to balloon-based observatories. Scientific ballooning is often used as step stone towards satellite missions, as for the UV MCP detector described in this publication. Following the first flight, it is foreseen to open STUDIO for the scientific community for use in future flights. The availability of a readily built telescope and gondola will greatly simplify the access to balloon-based science for astronomers.

The gondola is mostly constructed from commercially available parts. It accommodates a 50 cm aperture telescope optimized for the visible and UV range of the spectrum. A beam splitter on the optical bench separates the beams for the visible channel (330-1100 nm) and the UV channel (180-330 nm). An active two-stage image stabilization system will counteract movements of the gondola and the telescope to an accuracy of 0.5 arcsec on the imaging plane of the MCP detector. A more detailed description of ESBO *DS* and the STUDIO gondola and telescope can be found in Pahler et al. (2020)[23].

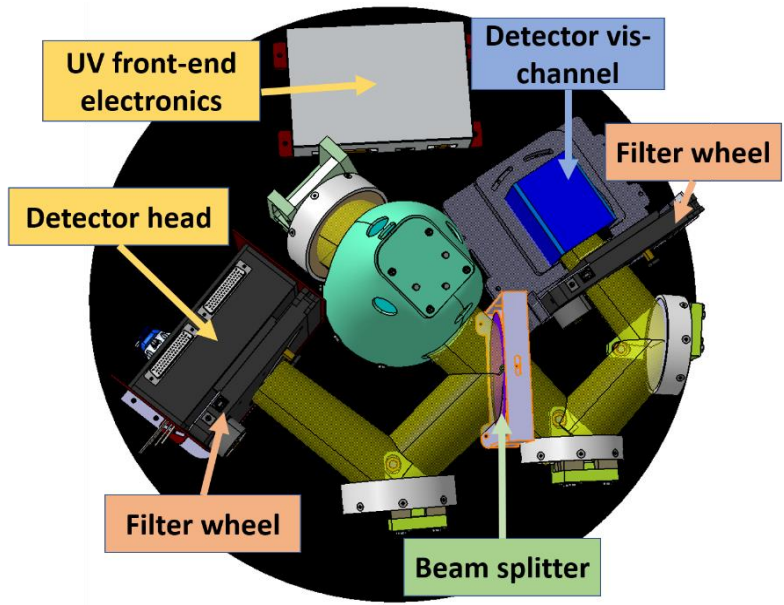

Figure 8. Telescope Instrument Platform of the STUDIO telescope

For the "Tübingen IIA Nebula Investigator" (TINI) space mission in collaboration with the Indian Institute of Astrophysics (IIA), we are developing an open FUV version of our detector. A lightweight detector door is developed by KTO (Kampf Telescope Optics GmbH, Munich). The CubeSat-sized instrument is capable of far-ultraviolet spectroscopic imaging in the wavelength range 92-180 nm with a spectral resolution of 0.1 nm and a spatial resolution of 7 arcsec.

With the Purple Mountain Observatory in China we are planning the "Census of WHIM Accretion Feedback Explorer" (CAFE) space mission. As a first step will demonstrate the capability of the instrument in the laboratory. It is a narrow band imager for O VI (1020-1080 Å) and H I (1240-1300 Å).

## 7. SUMMARY AND OUTLOOK

The space-qualified detector presented here can be optimized to the needs of different missions thanks to its compact size and low number of interfaces. Currently, we implement a centroiding algorithm in a custom FPGA board. After investigating the growth and the resulting QE of GaN on $MgF_2$, we aim at optimizing the gradient p-doping of the GaN layer to tweak the QE further. Alternatively, we have the option of a cesium telluride photocathode with a detection efficiency in semi-transparent mode of 3-5 %, which would translate to 6-10 % in opaque mode. For aligning and testing of the telescope instrument platform of the STUDIO instrument we are preparing a sealed detector body.

## ACKNOWLEDGEMENTS

The development of the IAAT UV MCP detector is funded by the *Bundesministerium für Wirtschaft und Technologie* through the *Deutsches Zentrum für Luft- und Raumfahrt e.V.* (DLR) under the grants 50 QT 1501 and 50 QT 2001.

ESBO *DS* has received funding from the European Union's Horizon 2020 research and innovation programme under grant agreement No 777516.